\documentclass[twocolumn,prb,superscriptaddress,preprintnumbers,amsmath,amssymb,showpacs,floatfix]{revtex4}

\usepackage{graphicx}
\usepackage{dcolumn}
\usepackage{bm}
\usepackage{color}
\usepackage{multirow}

\newcommand{\lco}{LaCoO$_3$}
\newcommand{\lsco}{La$_{2-x}$Sr$_x$CoO$_4$}

\newcommand{\tg}{$t_{2g}$}
\newcommand{\eg}{$e_{g}$}

\newcommand{\cha}{$\chi_{ab}$}
\newcommand{\chc}{$\chi_c$}
\newcommand{\ch}{susceptibility}

\newcommand{\coz}{Co$^{2+}$}
\newcommand{\cod}{Co$^{3+}$}

\begin{document}

\title{Evidence for a temperature-induced spin-state transition of Co$^{3+}$ in La$_{2-x}$Sr$_x$CoO$_4$}
\author{N.~Hollmann}
 \affiliation{II. Physikalisches Institut, Universit\"{a}t zu K\"{o}ln,
 Z\"{u}lpicher Str. 77, 50937 K\"{o}ln, Germany}
\author{M.~W.~Haverkort}
 \affiliation{Max-Planck-Institut f\"ur Festk\"orperforschung, Heisenbergstra\ss e 1, D-70569 Stuttgart, Germany}
\author{M.~Benomar}
 \affiliation{II. Physikalisches Institut, Universit\"{a}t zu K\"{o}ln,
 Z\"{u}lpicher Str. 77, 50937 K\"{o}ln, Germany}
\author{M.~Cwik}
 \affiliation{II. Physikalisches Institut, Universit\"{a}t zu K\"{o}ln,
 Z\"{u}lpicher Str. 77, 50937 K\"{o}ln, Germany}
\author{M.~Braden}
 \affiliation{II. Physikalisches Institut, Universit\"{a}t zu K\"{o}ln,
 Z\"{u}lpicher Str. 77, 50937 K\"{o}ln, Germany}
\author{T.~Lorenz}
 \affiliation{II. Physikalisches Institut, Universit\"{a}t zu K\"{o}ln,
 Z\"{u}lpicher Str. 77, 50937 K\"{o}ln, Germany}

\date{\today}


\begin{abstract}

We study the magnetic susceptibility of mixed-valent La$_{2-x}$Sr$_x$CoO$_4$ single crystals
in the doping range of $0.5\leq x \leq 0.8$ for temperatures up to 1000~K. The magnetism below room temperature is described by paramagnetic Co$^{2+}$ in the high-spin state and by Co$^{3+}$ in the non-magnetic low-spin state. Above room temperature, an increase in susceptibility compared to the behavior expected from Co$^{2+}$ is seen, which we attribute to a spin-state transition of Co$^{3+}$. The susceptibility is analyzed by comparison to full-multiplet calculations for the thermal population of the high- and intermediate-spin states of Co$^{3+}$. 

\end{abstract}

\maketitle

Cobaltates show a variety of unusual physical properties, among them 
unconventional superconductivity\cite{takada03a} and giant magneto resistance.\cite{perez98a} Moreover, the spin state of cobalt can act as an extra degree of freedom. Especially in oxides
containing octahedrally coordinated \cod, a competition may arise between the crystal-field splitting $10Dq$, which splits
the $3d$ orbitals into \eg\ and \tg\ states and the on-site Coulomb exchange or Hund's coupling. In a local view for \cod\ with a $3d^6$ configuration, a strong crystal field favors the non-magnetic low-spin state ($S=0$, $t_{2g}^{\,6}e_g^0$), while Coulomb exchange tends to stabilize the magnetic high-spin state ($S=2$, $t_{2g}^{\,4}e_g^2$). Apart from these two spin states, the intermediate-spin state ($S=1$, $t_{2g}^{\,5}e_g^1$) attracted also attention as band-structure effects\cite{korotin96a} and large distortions may stabilize this state.
In the rare case of a subtle balance between different spin states, a thermal population of different spin states becomes possible leading to unusual magnetic properties. Among the cobaltates, the pseudo-cubic perovskite LaCoO$_3$\cite{jonker53a,goodenough58a} is known to exhibit such a thermally-induced spin-state transition, having triggered a huge effort to understand its properties since more than 50 years. LaCoO$_3$ is in the low-spin state at low temperatures rendering the compound essentially non-magnetic. Up to 100~K, the \ch\ rises as another spin state is very close in energy to the low-spin state. The nature of the excited spin state that is populated in this spin-state transition is still under debate. Both experimental and theoretical studies claim to identify either the high-spin or the intermediate-spin state. For an overview on this topic see e.g. Ref.~\onlinecite{haverkort06a}, where x-ray absorption spectroscopy is used to identify the high-spin state as the excited state.

The layered perovskites \lsco\ are closely related to LaCoO$_3$ as they crystallize with a similar local coordination of the Co ions in oxygen octahedra. The layered cobaltates are mixed-valent, containing \coz\ and \cod\ in a ratio of ($1-x$):$x$, and the spin state of the cobalt ions is expected to play a decisive role in the electronic and magnetic properties. These materials are discussed in the recent literature.\cite{moritomo97a,itoh99a,zaliznyak00a, wang00a, wang00b, zaliznyak01a, sanchez04a, shimada06a, chichev06a, savici07a, ang08a, ang08b, sakiyama08a, hollmann08a, chang09a, cwik09a, helme09a, wu09a, wu10a, tealdi10a, babkevich10a} 
Many different scenarios for the spin state can be found in these publications: For example, the pure \cod\ compound LaSrCoO$_4$ ($x=1$) has been proposed to be in the intermediate-spin state\cite{chichev06a, moritomo97a, ang08a} and in a mixture of low-spin and high-spin states.\cite{wang00a, wu10a} Magnetic \ch\ data\cite{moritomo97a} and NMR results\cite{itoh99a} were interpreted with a transition from the high-spin ($x\leq 0.7$) to the intermediate-spin ($x>0.7$) state. Hartree-Fock calculations\cite{wang00a} yield the high-spin state for $x\leq 0.52$ and a low-spin/high-spin ordered phase for $x>0.52$, while in another publication\cite{zaliznyak01a} \cod\ in the $x=0.5$ compound is believed to be in the intermediate-spin state. In a previous paper,\cite{hollmann08a} we analyzed the magnetic \ch\ of a series of \lsco\ with $0.5\leq x \leq 0.8$ and used the magnetic \ch\ up to 400~K and its anisotropy to deduce the spin state of \cod. Both the anisotropy and the general behaviour of the \ch\ in the paramagnetic case could be described by taking only \coz\ in the high-spin state into account. Consequently, the ground state of \cod\ was proposed as the low-spin state. This arrangement of spin states for \coz\ (high spin) and \cod\ (low spin) was confirmed by x-ray absorption experiments\cite{chang09a} and it was also shown that the spin state drastically affects electron hopping through a spin-blockade mechanism, as the high-spin state of \coz\ and the low-spin state of \cod\ are not preserved under electron hopping. Neutron scattering revealed for $x\simeq 0.4-0.6$ stripe ordering patterns between the magnetic \coz\ and the non-magnetic \cod\ ions.\cite{cwik09a} Theoretical support for this spin-state scenario of the $x=0.5$ compound comes from local-density approximation calculations.\cite{wu09a}

Above room temperature, some deviations from the model with non-magnetic \cod\ appeared in the analysis of the \ch\ of \lsco,\cite{hollmann08a} where an additonal moment starts to rise in the experimental curves.
In the present paper we analyze high-temperature \ch\ measurements on \lsco\ up to 1000~K, to study the possiblity of a spin-state transition from the low-spin ground state to a higher spin state. We find that the rise in \ch\ at high temperatures is indeed connected to the thermal population of a higher spin state, thus identifying a temperature-induced spin-state transition of \cod\ in \lsco.  

\begin{figure}[t]
\includegraphics[angle=0,width=9cm]{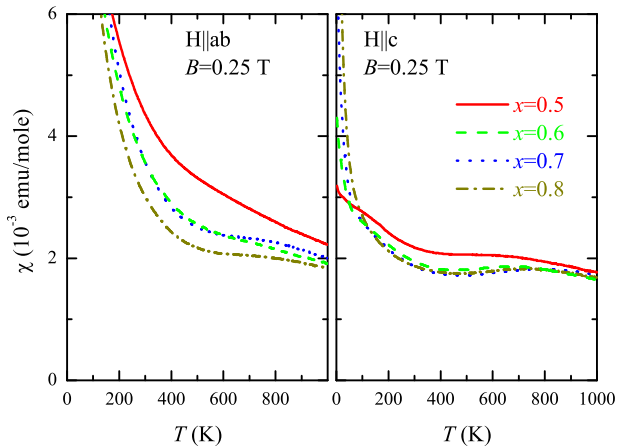}
 \caption[]{(color online) Magnetic \ch\ of \lsco\ for two different directions of the magnetic
 field. The left panel displays the \ch\ for the field perpendicular to the tetragonal $c$ axis, the right panel for a field parallel to the $c$ axis. The anisotropy is of easy-plane type up to 1000~K.} \label{fig:chi}
\end{figure}

The \ch\ measurements were carried out with a vibrating sample magnetometer (VSM) in a Physical Properties Measurement System of Quantum Design equipped with an oven insert for a temperature range of 300~K $\le T\le$ 1000~K. The low-temperature data recorded previously\cite{hollmann08a} had been taken up to 400~K, and the overlap of the two data sets between 300~K and 400~K is satisfactory. The curves are identical in this range after scaling by small factors in the order of a few percent. For each sample, the heating cycle was measured twice to ensure reproducibility of the curves, and to exclude a possible degradation of the samples at higher temperatures. An additional thermal gravimetric analysis (TGA) carried out in N$_2$ atmosphere shows that the mass deviation up to 1000~K is less than 0.1\%. This means that the sample composition and the stoichiometry remain stable at these high temperatures.

The magnetic \ch\ is shown in Fig.~\ref{fig:chi}, with the field applied perpendicular (left panel) and parallel to the tetragonal $c$ axis (right panel). The anisotropy observed is of easy-plane type \cha$>$\chc. Regarding the paramagnetic phase above $\sim 100$~K, \chc\ is rather similar for all doping levels $x$, while \cha\ has the tendency to become smaller with increasing $x$. As discussed in detail in Ref.~\onlinecite{hollmann08a}, the form of the curves at lower temperatures
is determined by the single-ion anisotropy of \coz. The \ch\ of \coz\ follows a complex paramagnetic behaviour due to crystal-field effects up to 200~K, which turns over into a Curie-Weiss-like susceptibility only for higher temperatures, where it scales roughly with the inverse of the temperature. 

\begin{figure}[t]
\includegraphics[angle=0,width=8cm]{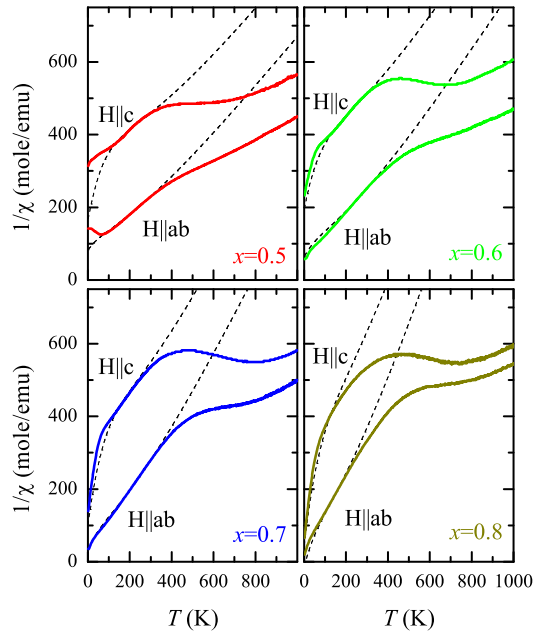}
 \caption[]{(color online) Inverse \ch\ of \lsco, drawn as solid lines, showing the non Curie-Weiss characteristics and the deviations from the expected behaviour of \coz, drawn as dashed lines.} \label{fig:invcalc}
\end{figure}

Above the sharp drop or Curie-like tail of the \ch\ from low temperatures up to 400~K, another effect is observed. The \ch\ curves exhibit a plateau, indicated by a reduction of the slope in \cha\ for small $x$ to an almost constant \ch\ for higher $x$. The effect can be even better seen in \chc, with an increase in \ch\ up to a broad maximum around 700-800~K for all $x$. Above these temperatures, the curves drop again with rising temperature. These plateaus indicate an additional source of magnetic moment that is induced by temperature, which we discuss in the inverse \ch.

The inverse \ch\ is shown in Fig.~\ref{fig:invcalc} as solid lines. In an inverse plot, the deviations from Curie-Weiss behaviour can directly be seen. The curves are only linear in a small temperature range of about 200~K to 300~K. At higher temperatures the inverse plots are bent downwards, indicating the occurence of higher momenta at higher temperatures. We will compare the inverse \ch\ for the whole temperature range with the full-multiplet calculations for \coz, which were used to verify the direction of the single-ion anisotropy and to explain the general behaviour of the curves at lower temperatures.\cite{hollmann08a} The \ch\ was calculated by diagonalizing a Hamiltonian for the Co $3d$ shell, including spin-orbit coupling, on-site Coulomb interaction, the external magnetic field, a mean exchange field, and an ionic tetragonal crystal field. The calculated inverse \ch\ curves are plotted as dashed lines in Fig.~\ref{fig:invcalc}. Apart from the low-temperature range of the half-doped ($x=0.5$) compound, which orders magnetically around 40~K,\cite{zaliznyak00a} the low-temperature \ch\ is described correctly within the model. Differences between the model and experimental data at temperatures above $\sim 300$~K are clearly visible. Note that deviations at higher temperatures with lower absolute values appear exaggerated in an inverse plot. The inverse \ch\ becoming lower than expected means that an additional moment is created in the experiment. \coz\ is not expected to show any features at these high temperatures. The \ch\ of \coz\ is given by the thermal population of low-lying local crystal-field and spin-orbit coupling excitations; they are in the order of few hundreds of Kelvin, and produce the change in slope at 50-100~K in the curves. At higher temperatures the \ch\ is expected to roughly follow a Curie-Weiss-like form, being almost linear in the inverse plot (Fig.~\ref{fig:invcalc}, dashed lines). The rise in \ch\ cannot be explained by a spin-state transition of \coz, as it is already in the high-spin state at lower temperatures; a spin-state transition to the low-spin state of \coz\ ($S=1/2$, $t_{2g}^{\,6}e_g^1$) would reduce the magnetic moment. The maximum of the \ch\ cannot be due to two-dimensional exchange coupling between the \coz\ ions, because the coupling itself is too weak\cite{cwik09a,helme09a,cwik} to produce maxima in the \ch\ at such high temperatures. Moreover, with increasing $x$ the maxima would shift to lower temperatures due to the decreasing \coz\ content, which is not observed in the experiment. Finally, the low-temperature magnetic \ch\ would not show an upturn in \ch\ at lower temperatures, as it is the case here. Thus, we identify \cod\ as the source of the additional moment. This is also in accordance with the doping dependence: the increase in \ch\ gets more pronounced for higher $x$, meaning more \cod. Because \cod\ is in the non-magnetic low-spin state at low temperatures, the only possibility to generate magnetic moment at the \cod\ site is a thermally induced spin-state transition to the high-spin or intermediate-spin state.
\begin{figure}[t!]
\includegraphics[angle=0,width=8.5cm]{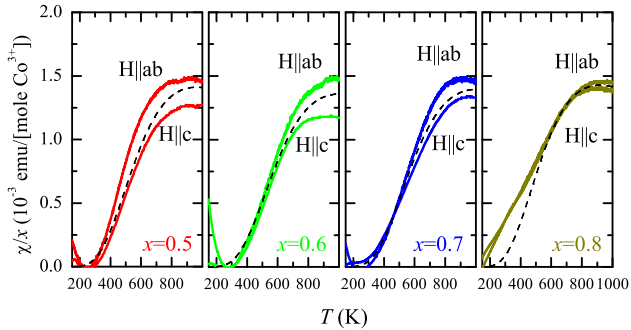}
 \caption[]{(color online) Susceptibility of \lsco\ after subtraction of the calculated curves for \coz, normalized by the \cod\ content $x$. The remaining \ch, plotted with solid lines, indicates the spin-state transition of \cod. The dashed lines are calculated by using a full-multiplet model for the spin-state transition to the high-spin state.} \label{fig:subtract}
\end{figure}

Next we study the amount of moment that is created by \cod\ at high temperatures and the characteristics of the excited spin state. For this quantitative analysis we first subtract the calculated \ch\ of \coz\ from the measured curves.\cite{subremark} The remaining moment is then identified with \cod\ (Fig.~\ref{fig:subtract}, solid lines). The total moment increases with the \cod\ content, and we have shown the \ch\ normalized by the \cod\ content $x$. It can be seen that the moment scales consistently with the doping. The anisotropy of the curves is small, and the uncertainty due to the subtraction is too large to use the anisotropy for an analysis within tetragonal symmetry. 

Using full-multiplet calculations, we can calculate the magnetic \ch\ of a spin-state transition to either the high-spin or the intermediate-spin states. Let us first discuss the high-spin state. 
\begin{figure}[b]
\includegraphics[angle=0,width=8cm]{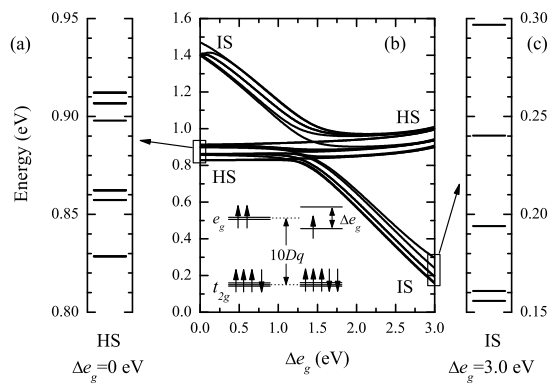}
 \caption[]{Energy levels of (a, HS) high- and (c, IS) intermediate-spin state; the small splittings are caused by spin-orbit coupling effects. The energies are given relative to the ground state (the low-spin state). (b) The intermediate-spin state can be stabilized by a large energy difference of the two \eg\ orbitals. A sketch of the orbital occupation for the two spin states is shown.} \label{fig:Elevel}
\end{figure}
By approximating the crystal field as cubic with a splitting of $10Dq\gtrsim 2.7$~eV, the high-spin state is the first excited state above the low-spin ground state. The energy gap between low and high-spin state depends on the relation of on-site Coulomb interaction, given by the Slater integrals, and the cubic crystal field splitting $10Dq$. Spin-orbit coupling is included in the calculations and affects the high-spin state that has a pseudo-orbital momentum of $\tilde{L}=1$ and spin $S=2$. As shown in Fig.~\ref{fig:Elevel}~(a), the 15-fold degenerate high-spin state is split into several states, which originate from a low-lying triplet with $\tilde{J}=1$, a quintet with $\tilde{J}=2$ higher in energy, and a septet with $\tilde{J}=3$ above the quintet. With the thermal population of the high-spin state all of these spin-orbit-split multiplets are populated.

To find a theoretical description for the experimental data, we keep the Slater integrals and spin-orbit coupling constant fixed.\cite{param} We only vary the cubic crystal field splitting $10Dq$, which determines the energy difference of low- and high-spin states that is the energy gap being responsible for the onset of moment for \cod. 
The \ch\ at high temperatures is overestimated within the model of the transition to the high-spin state and we apply a factor $f$ to reduce the moment of the spin-state transition. 
The reduction of moment for the high-spin state could be explained by covalency effects, by a temperature dependence in the energy gap like it was proposed for LaCoO$_3$,\cite{haverkort06a} or by antiferromagnetic exchange between \coz\ and \cod\ moments. The dashed lines in Fig.~\ref{fig:subtract} represent the calculated curves for the \ch.\cite{vv} The curves are described by the parameters $10Dq$ and $f$ listed in Table~\ref{tab:cod}. In the Table we also specify the resulting energy gap $\Delta$ in Kelvins, which defines the difference in energy between the low-spin ground state and the lowest level of the high-spin state, the $\tilde{J}=1$ triplet.

\begin{table}

\caption{\label{tab:cod} Values for the full-multiplet calculations for \cod\ for high-spin (HS) and intermediate-spin (IS) state transitions. $10Dq$ is the cubic crystal field splitting that is used as free parameter; $f$ is the prefactor scaling the moment. $\Delta$ is the resulting energy gap between the low-spin state and the high- or intermediate-spin state.}

\begin{ruledtabular}
\begin{tabular}{cccccc}

 & $x$ & 0.5 & 0.6 & 0.7 & 0.8\\
\hline
HS & $f$ & 0.73 & 0.74 & 0.76 & 0.70 \\
 & $10Dq$ (eV) & 2.805 & 2.810 & 2.810 & 2.800 \\
 & $\Delta$ (K) & 1960 & 2060 & 2060 & 1850 \\
\hline
IS & $f$ & 1.70 & 1.80 & 1.80 & 1.70 \\
$\Delta e_g$=3.0 eV & $10Dq$ (eV) & 3.150 & 3.160 & 3.160 & 3.140\\
 & $\Delta$ (K) & 1810 & 1900 & 1900 & 1720 \\

\end{tabular}
\end{ruledtabular}
\end{table}

In a cubic crystal field in such a local picture, the intermediate-spin state cannot be the first excited state above the low-spin state. It can, however, be stabilized by splitting the $e_g$ orbitals in energy. Such a splitting can be caused by the local distortions of the CoO$_6$ octahedra which are possible in this layered perovskite structure, where the Co-O bond lengths are larger out-of-plane than in-plane. In order to model the \ch\ we choose a splitting of $\Delta e_g=3.0$~eV that lowers the energy of the $d_{3z^2-r^2}$ orbital with respect to $d_{x^2-y^2}$. This splitting is far away from the crossing of high- and intermediate-spin states, see Fig.~\ref{fig:subtract}~(b). As for the high-spin state, the crystal field splitting $10Dq$ is tuned to find a description of the experimental data. The \ch\ curves calculated for the intermediate-spin state are not shown here, because they are practically identical to the curves for the high-spin state shown as dashed lines in Fig.~\ref{fig:subtract}. The quality of the fit for the model of the intermediate-spin state is thus similar to the model of the high-spin state. The parameters used for the intermediate-spin state are also listed in Table~\ref{tab:cod}. The energy gap $\Delta$ is slightly smaller than that for the high-spin state, and the calculated moment of the intermediate-spin state is too small compared to the moment observed, which is corrected by a factor $f$ larger than one. This factor could be explained by ferromagnetic exchange between the magnetic moments of \coz\ and \cod.

The parameters in Table~\ref{tab:cod} for all dopings are fairly similar, showing that the moment created in the spin-state transition scales correctly with the \cod\ content and that the energy gap between the spin states is almost independent of doping. This agrees with the small structural changes\cite{cwik} in the $x$ range that is studied here. The shape of the \ch\ is also described correctly by both high- and intermediate-spin states, except for $x=0.8$ where the \ch\ in the temperature range 200 to 400~K is underestimated. Here, the description of the \ch\ of \coz\ in Fig.~\ref{fig:invcalc} might be not accurate enough. We also could speculate that the early onset of moment is a sign of a smaller energy gap of this compound. This would mean that the higher-spin state of \cod\ becomes more stable with higher $x$ and would fit to the theoretical expectation\cite{wu10a} that half of the \cod\ ions should be in the high-spin state for the $x=1$ compound. 

Overall, the full-multiplet model for \cod\ gives a reasonable description of the shape of the \ch\ at higher temperatures; for a clear distinction between high- and intermediate-spin state the analysis of the magnetic \ch\ is not sufficient. To reproduce the moment observed at the spin-state transition, the calculations are scaled with the prefactor $f$, which results in a reduction of moment for the high-spin state and an increase for the intermediate-spin state. In principle, it is also possible to describe the data without the prefactor $f$ by assuming an excited spin state with a mixed character of high and intermediate spin. Such a state can be found in calculations with an energy splitting of $\Delta e_g\approx 2.3$~eV, see Fig.~\ref{fig:subtract}~(b). It is, however, doubtful whether such an analysis is useful, as besides the experimental errors caused by the subtraction of the \ch\ of \coz, the model itself is probably too simple: covalency and band formation, as well as magnetic exchange are neglected in the local crystal field picture. We therefore propose to clarify the nature of the excited spin state of \cod\ by means of other methods e.g., x-ray absorption spectroscopy and neutron scattering. 

The energy gap that determines the onset of the moment from the spin-state transition is fairly independent from the nature of the excited spin state, and thus it can be deduced from the \ch\ measurements. We compare the energy gap $\Delta$ to other cobaltates that show a thermally-induced spin-state transition. The values of $\Delta$ in the layered cobaltates \lsco\ are higher than for the case of \lco, where the \ch\ already has its maximum around 100~K with $\Delta\simeq 185-256$~K.\cite{zobel02a} The Co-O bond in \lco\ is 1.91~\AA, while the in-plane bond length for \cod\ in La$_{1.5}$Sr$_{0.5}$CoO$_4$ is smaller (1.89~\AA),\cite{cwik} effectively enhancing the crystal field in the latter compound. This stabilizes the low-spin ground state and pushes the excited spin states to higher energies. The $\Delta$ in \lsco\ is still smaller than in EuCoO$_3$, where 3000~K is found,\cite{baier05a} which results from the strong compression of the octahedra due to the small size of the Eu$^{3+}$ ions.

In conclusion, we have studied the high-temperature \ch\ of \lsco. We find that the \cod\ ions, which are in the low-spin state at low temperatures, show a thermal population of a higher spin state. The experimental data are well described by a full-multiplet model for a spin-state transition, yielding an energy gap of around 2000~K between the spin states. The results suggest to investigate this spin-state transition further by other experimental methods in order to deduce the nature of the excited spin state.

We would like to thank S.~Heijligen for carrying out the TGA experiment and additional SQUID measurements. Financial support by the Deutsche
Forschungsgemeinschaft through SFB~608 is acknowledged. N.~H. is also funded by the Bonn-Cologne Graduate School.

\bibliographystyle{apsrev}


\begin{thebibliography}{30}
\bibitem{subremark}{Subtracting a Curie-Weiss law from the curves instead of the calculations for \coz\ yields a similar result for $0.5\leq x\leq 0.7$. The inverse \ch\ of $x=0.8$ is not linear in any temperature range, and thus a Curie-Weiss fit is not applicable.}
\bibitem{param}{The values have been taken from Hartree-Fock calculations for \cod. Slater integrals: $F_2=12.662$~eV and $F_4=7.916$~eV, single-electron spin-orbit coupling constant: $\zeta=74$~meV.}
\bibitem{vv}{A small constant van-Vleck contribution is subtracted from the calculations.}
\end{thebibliography}

\end{document}